\documentclass[aps,superscriptaddress,twocolumn,superscriptaddress,showkeys,groupedaddress,nofootinbib,nobalancelastpage,nobibnotes]{revtex4}
\pdfoutput=1
\usepackage{amsmath,amsfonts,amssymb,mathrsfs,graphicx,color,epsfig}
\usepackage[squaren]{SIunits}
\usepackage{verbatim}
\usepackage{enumerate}
\usepackage[hyperfootnotes=false]{hyperref}
\usepackage{xcolor}
\usepackage{slashed}
\usepackage[utf8]{inputenc}
\usepackage{rotating}
\usepackage{isotope}

\newcommand{\ie}{{\it i.e.}}

\newcommand{\eg}{{\it e.g.}}

\newcommand{\eq}{Eq.}

\newcommand{\fig}{Fig.}

\newcommand{\Ref}{Ref.}
\newcommand{\Refs}{Refs.}

\newcommand{\thetaobs}{\theta_\text{obs}}
\newcommand{\thetajet}{\theta_\text{jet}}

\newcommand{\equ}[1]{\eq~(\ref{equ:#1})}
\newcommand{\figu}[1]{\fig~\ref{fig:#1}}

\newcommand{\bi}{\begin{itemize}}
\newcommand{\ei}{\end{itemize}}

\begin{document}


\title{Expected neutrino fluence from short Gamma-Ray Burst 170817A, \\ and off-axis angle constraints}

\author{Daniel Biehl}
\affiliation{Deutsches Elektronen-Synchrotron (DESY), Platanenallee 6, D-15738 Zeuthen, Germany}

\author{Jonas Heinze}
\affiliation{Deutsches Elektronen-Synchrotron (DESY), Platanenallee 6, D-15738 Zeuthen, Germany}

\author{Walter Winter}
\affiliation{Deutsches Elektronen-Synchrotron (DESY), Platanenallee 6, D-15738 Zeuthen, Germany}

\date{\today}

\begin{abstract}
\vspace{0.5cm}
We compute the expected neutrino fluence from SGRB 170817A, associated with the gravitational wave event GW 170817, directly based on Fermi observations in two scenarios:  structured jet and off-axis (observed) top-hat jet. 
While the expected neutrino fluence for the structured jet case is very small, large off-axis angles  imply high radiation densities in the jet, which can enhance the neutrino production efficiency. In the most optimistic allowed scenario, the neutrino fluence can reach only $10^{-4}$ of the sensitivity of the neutrino telescopes.
We furthermore demonstrate that the fact that gamma-rays can escape limits the baryonic loading (energy in protons versus photons) and the off-axis angle for the internal shock scenario. In particular, for a baryonic loading of ten, the off-axis angle is more strongly constrained  by the baryonic loading than by the time delay between the gravitational wave event and the onset of the gamma-ray emission. 
\end{abstract}


\maketitle

\section{Introduction}

The gravitational wave event GW170817~\cite{TheLIGOScientific:2017qsa} has recently been drawing a lot of attention because it was accompanied by electromagnetic counterparts, first in gamma-rays~\cite{Monitor:2017mdv}, and later also in the X-ray~\cite{Troja:2017nqp,Margutti:2017cjl} and radio~\cite{Hallinan:2017woc,Alexander:2017aly} bands. 

The UV, optical, and near-infrared observations have been interpreted as kilonova~\cite{Smartt:2017fuw,Nicholl:2017ahq,Cowperthwaite:2017dyu}, with evidence for the synthesis of heavy r-process elements~\cite{Kasen:2017sxr,Chornock:2017sdf}. A detailed overview of the many multi-wavelength and multi-messenger observations can be found in \Ref~\cite{GBM:2017lvd} and references therein; for a theoretical interpretational overview, see \eg\ \Ref~\cite{Metzger:2017wot}.

There has been a follow-up study searching for high-energy neutrinos in a wide energy range (100~GeV to 100~EeV) by the ANTARES, IceCube and Auger collaborations~\cite{ANTARES:2017bia}, finding nothing. The results have been interpreted in terms of models predicting neutrinos from neutron star-neutron star mergers~\cite{Kimura:2017kan,Fang:2017tla}, re-scaled to the observed distance. However, no direct prediction for this short Gamma-Ray Burst (SGRB) event 170818A has been derived. 

In this study, we use the information on the spectral energy distribution (SED) from {\it Fermi}-GBM~\cite{Goldstein:2017mmi} together with the inferred parameters of the SGRB jet directly to predict the expected neutrino fluence for this event. We consider two different jet models, a structured jet and a uniform top-hat jet observed off-axis. We also discuss the relevance of the photospheric constraint in the off-axis scenario.

\section{Considered jet models}

\begin{figure*}[t!]
\centering
\includegraphics[width=0.7\textwidth]{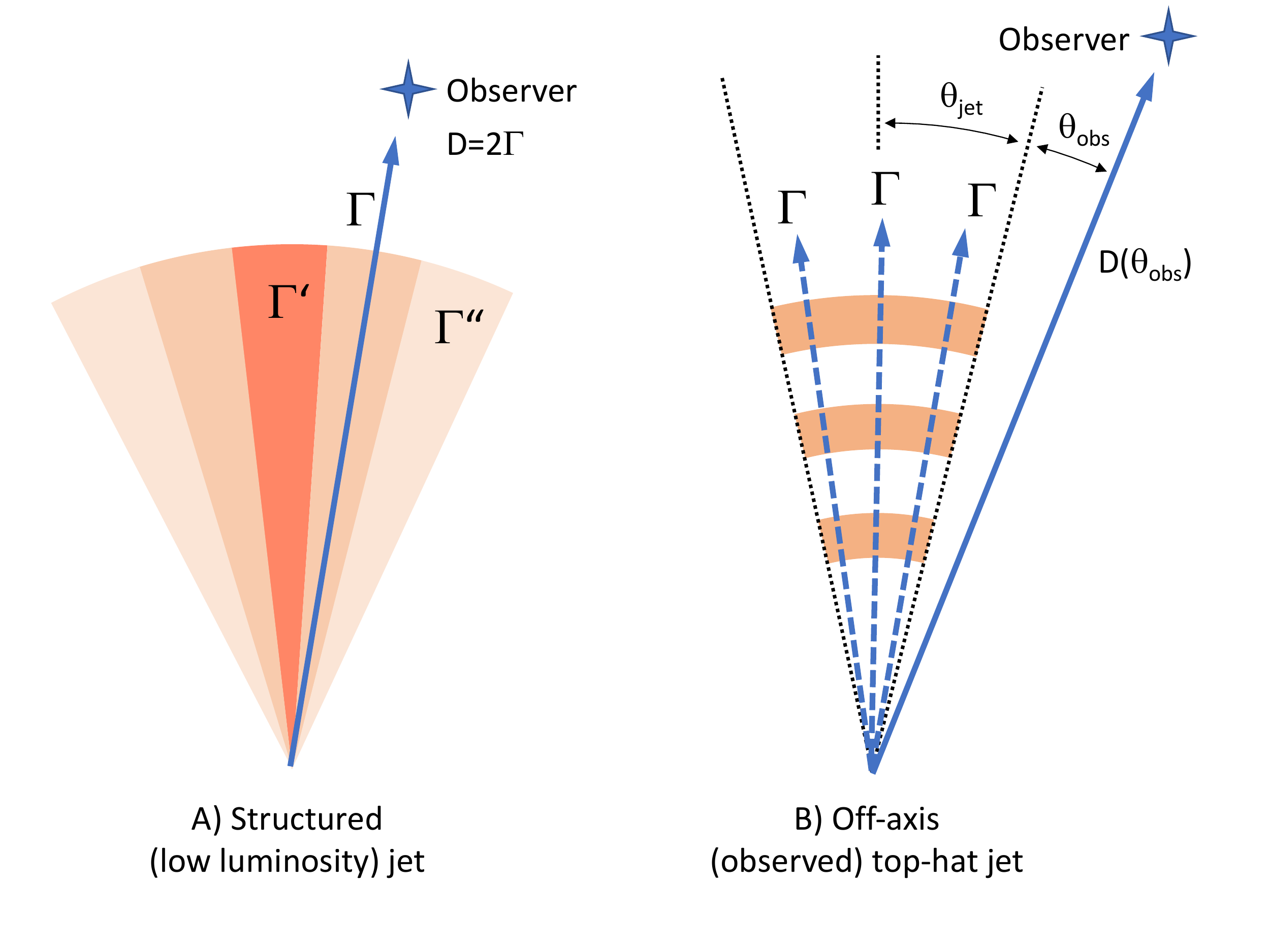}
\caption{\label{fig:jetgeo}Illustration of different jet geometries and corresponding angles used in this work.} 
\end{figure*}

Predicting the neutrino production for this event depends on the jet model as the observed emission has to be boosted back into the jet frame where the interactions take place. 
We consider two of the models given in \Ref~\cite{Monitor:2017mdv} to explain the low luminosity of GRB170817A: A structured jet and a uniform top-hat jet observed off-axis (similar to the simple and advanced models in~\cite{Denton:2017jwk}) -- as illustrated in \figu{jetgeo}.

Note that we clearly distinguish between $\Gamma$-factor of the jet and the Doppler factor $D$ for the boost from the jet rest frame to the observer's frame in this work. The Doppler factor is given as a function of off-axis angle $\theta$ and Lorentz factor $\Gamma$ by
\begin{align}
  D(\theta) = \frac{1}{\Gamma (1- \beta \cos\theta)} \approx \frac{2 \Gamma}{1 + \theta^2 \Gamma^2} .
\label{equ:D}
\end{align}
For a jet observed on-axis this means $D(0) = 2 \Gamma$; the effect of the cosmological redshift can be neglected because of the small redshift under which GW170817 has been observed.

In the  structured jet scenario \figu{jetgeo},~A) the jet is structured in the sense that has different characteristics ($E$, $\Gamma$) if observed under different angles. If we observe the structured jet at an axis that has low luminosity, this naturally explains the dimness of GRB170817A.
Note that the observer will also be exposed to the off-axis emission from the other axes of the jet (illustrated with $\Gamma'$ and $\Gamma''$ in the figure). Though they are suppressed due to their Doppler factor, they might overshoot the on-axis radiation if the gradient in luminosity is too high. This implies that the $\Gamma$-factor and brightness must not change too rapidly over the viewing angle. As long as these variations are small, the structured jet can  be treated as a low luminosity top-hat jet observed on-axis. We will refer to this case as ``low luminosity'' jet. Its predictions are close to the conventional GRB neutrino flux predictions scaled to low luminosity, see \eg\ \Refs~\cite{Waxman:1997ti,Guetta:2003wi,Hummer:2011ms,He:2012tq,Tamborra:2015qza}.
Note that a trivial version of the structured jet is a uniform emission into all directions. This case can be  motivated by postulating that observing GRB170817A in coincidence with the gravitational wave event is more than a coincidence -- whereas the probability to actually sit close enough to the jet axis for an arbitrary neutron star merger is relatively small. Other alternatives, such as a cocoon emission, are not treated here; see also \Ref~\cite{Mooley:2017enz} for a very recent discussion of possible geometries.

In the off-axis scenario \figu{jetgeo},~B) we assume a uniform top-hat jet that is observed off-axis. For an off-axis observer the luminosity of the jet is suppressed due to the Doppler factor, which depends on the angle between the edge of the jet and the observation axis $\thetaobs$; see below 
for details on these transformations.
We do not impose any direct constraint on $\theta_{\mathrm{jet}}$, but we will see a transition in the luminosity scaling around $\theta_{\mathrm{obs}} \simeq \theta_{\mathrm{jet}}$ due to the observed geometry. While the jet-opening angle $\thetajet$ is not inferable from observations, we use the estimate $\theta_{\mathrm{jet}} \simeq 1/\Gamma$ to demonstrate this transition.

In both cases we assume the usual geometry of a relativistically expanding fireball. In this scenario, internal shocks are created in the collision of coasting plasma shells with Lorentz factor $\Gamma$ (see \eg\ \cite{Kobayashi:1997jk}) which are driven into the environment of the central engine in all directions within the jet opening angle $\theta_{\mathrm{jet}}$.
For an on-axis observer, this fireball is not distinguishable from a sphere expanding into the full solid angle, which allows to treat the fireball in the isotropic-equivalent picture. 
For the low luminosity jet we can take the observed quantities at face value, while for the off-axis scenario, we have to transform the observation to the on-axis frame as a function of $(\thetaobs,\Gamma)$. For $\thetaobs = 0^\circ$, both scenarios coincide by construction.

Concerning the neutrino production in the off-axis fireball scenario, we note that the common intuition of lower expected neutrino fluxes for larger off-axis angles (as \eg\ in \Ref~\cite{ANTARES:2017bia}) is wrong if the observed gamma-ray fluence is fixed. Since the observed gamma-ray fluence has to be de-boosted by the Doppler factor, the photon density in the jet frame will be much higher compared to the structured jet case, and as a consequence the neutrino production efficiency; the boost back into the observer's frame cannot compensate for that, which means that the expected neutrino flux will be higher for large off-axis angles due to this simple re-scaling.

\section{Methods and Assumptions}

\subsection{{\it Fermi} observations}
\label{sec:observations}

{\it Fermi}-GBM measured the duration of the burst $T_{90} = 2.0 \pm 0.5$~s and the minimum variability timescale $t_v = 0.125 \pm 0.064$~s \cite{Goldstein:2017mmi}. With the accompanying gravitational wave signal, the redshift of the source was determined to be $z = 0.008^{+0.002}_{-0.003}$ \cite{GBM:2017lvd}. Together with the gamma ray flux $F_\gamma = (5.5 \pm 1.2) \times 10^{-7}$ erg s$^{-1}$ cm$^{-2}$ measured by {\it Fermi}-GBM \cite{Goldstein:2017mmi}, this gives an estimate of the luminosity of the source $L_\gamma = 10^{46.9}$ erg s$^{-1}$, and a corresponding $E_\gamma = 1.6 \cdot 10^{47}$ erg -- which is in the same range as given in \Ref~\cite{Monitor:2017mdv}.

Several approaches to fit the spectral energy distribution (SED) of this event have been presented by the Fermi collaboration. Here, we follow the description of the SED as a comptonized spectrum, corresponding to the best-fit to the observation reported in \Ref~\cite{Monitor:2017mdv}.
We use the 256 ms time-integrated selection from T0-0.192s to T0+0.064s, for which the fit yields a spectral index of $\alpha = 0.14 \pm 0.59$ and a peak energy $E_\text{peak} = 215 \pm 54$ keV \cite{Goldstein:2017mmi}. Note that the results in the following sections will slightly change depending on the time interval and energy band chosen to determine the parameters.
So far, there is no conclusive information about the Lorentz factor $\Gamma$ and the observation angle $\theta_\text{obs}$. They can however be constrained by the time delay between gravitational wave and electromagnetic signal \cite{Granot:2017gwa,Salafia:2017hfr}. In the following, we will compare our results  to this constraint adopted to our geometry.

\subsection{Dissipation radius}

The critical input parameters for the neutrino production efficiency are the gamma-ray (isotropic equivalent) luminosity $L_\gamma$  and the dissipation radius $R_\text{diss}$, \ie, the distance of the production region from the central engine. In the internal shock scenario, $R_\text{diss}$ is equal to the collision radius  $R_\text{coll}$ where the shells in the jet collide, shocks form, and particles are accelerated. It is usually estimated by the relationship
\begin{align}
  \label{equ:collision-radius}
  R_\text{coll} \simeq 2\Gamma^2 c t_v 
\end{align}
from the Lorentz factor $\Gamma$ and the variability timescale $t_v$ in the source frame. Note that $R_\text{coll}$ and $\Gamma$ are both given in the source frame (and therefore independent on the off-axis angle), while $t_v$ is given by the on-axis observation.

Radiation from internal shocks can only be directly observed if $R_\text{coll} \gtrsim R_\text{ph}$, where the photospheric radius $R_\text{ph}$ is in this work defined as the radius where the shells become optically thin to Thomson scattering.
The photospheric radius is, for our geometry (the coasting phase of the shells~\cite{Daigne:2002zg}), given by
\begin{align}
  \label{equ:photospheric-radius}
  R_\text{ph} & \simeq  \left(\frac{\sigma_T}{4 \pi m_p} \right)^{1/2} \, \left(\frac{\xi_A}{\varepsilon} \frac{E_\text{iso,on}}{\Gamma T_{90}/t_v} \right)^{1/2} \, ,
\end{align}
where $E_\text{iso,on}$ is the total isotropic equivalent energy in $\gamma$-rays for an on-axis observer, $\varepsilon$ is the conversion efficiency of kinetic energy to total dissipated energy, and $\xi_A$ is the baryonic loading defined as ratio between energy in protons and photons in the {\it Fermi}-GBM energy band from 10 to 1000 keV.\footnote{To derive this formula we assume that the optical depth of the shell to Thomson scattering is determined by  thermal electrons $\tau_T^\prime = n^\prime_e \sigma_T d^\prime \simeq 1$ ($d'$: shell width, $n^\prime_e$: electron density). The total number of thermal electrons is estimated from the number of protons using charge conservation, assuming negligible contribution from electron-positron pair production: $N_e \sim  N_p \simeq M_{\text{shell}}/m_p = E_{\text{kin}}/(\Gamma m_p)$. Here it is implied that the bulk kinetic energy of the shell is dominated by baryons; it can be estimated from the dissipated energy $E_{\text{diss}} \simeq \xi_A E_{\gamma,\mathrm{iso}}$ if the  dissipation efficiency $\varepsilon \equiv E_\text{diss}/E_\text{kin}$ is known.} 
In the following we use $\varepsilon = 25 \%$ as an estimate corresponding to the values found in \Ref~\cite{Bustamante:2016wpu}, see also \Ref~\cite{Beloborodov:2000nn}. Note that too small values are not compatible with the energy observed in the afterglow, whereas much higher efficiencies are difficult to obtain in these models because they require collisions of shells with extremely different Lorentz factors. 
Furthermore, note that the photospheric radius scales with the ratio $\xi_A / \varepsilon$, which means that lower values for $\varepsilon$ corresponds to a lower baryonic loading for fixed $R_{\mathrm{ph}}$. Demanding that $R_\text{coll} \gtrsim R_\text{ph}$ will later  constrain the baryonic loading in the internal shock scenario.

 It is possible to derive an estimate for the maximal $R_{\mathrm{diss}}$  for  SGRB170817A  from the delay time $t_\text{delay}$ between the gravitational wave and electromagnetic signal~\cite{Monitor:2017mdv}. Assume that the emission originates from two colliding shells with Lorentz factors $\Gamma_1$ and $\Gamma_2$ with $\Gamma_2 > \Gamma_1$. If the first shell is emitted at the time of the merging, the distance it has traveled by the time the second shell catches up with it will be $R_\text{coll} \approx 2\Gamma_1^2 c t_\text{delay}$. Thus, for $\Gamma_1 < 100$, the upper limit on the collision radius is around $R \sim 10^{9.5}$ km. In our model, we typically deal with dissipation radii between $10^7$ and $10^8$ km, which are well below this limit and thus consistent with it. 

\subsection{On-off-axis transformations}

For an emitting shell moving at relativistic speed, the observed quantities such as energy and time will be Doppler shifted depending on the observation angle and Lorentz factor of the emitting shell
\begin{align}
  t &= D(\thetaobs)^{-1} \,t^\prime \\
  E &= D(\thetaobs) \, E^\prime
\end{align}
with the Doppler factor in \equ{D}.
These transformations are valid for quantities such as the peak energy $E_{\gamma,\text{peak}}$, which can be defined in the shell frame.
They are however not necessarily valid for observed quantities that have to be integrated over the geometry of a single shell, \ie\ for the isotropic equivalent energy $E_\text{iso}$ and variability time $t_v$, and for the duration $T_{90}$ which is only defined for the whole burst.

The isotropic equivalent energy $E_\text{iso}$ does not scale as one would naively expect for an energy. This is because it is defined as the observed spectral flux $F_\nu$ in $[\text{erg} \, \text{s}^{-1} \, \text{cm}^{-2} \, \text{Hz}^{-1}]$ integrated over time, area and frequency. It therefore scales differently depending on whether the observer is inside or outside of $\theta_\text{jet}$. A full derivation can be found in the Appendix of \Ref~\cite{Ioka:2017nzl}, arriving at:
\begin{align}
\nonumber
  E_\text{iso}(\thetaobs) &\propto 
  \begin{cases}
    \text{const}
    &\text{for } \thetaobs \lesssim 0 \\
    D(\thetaobs)^2
    &\text{for } 0 < \thetaobs \lesssim \thetajet \\
    D(\thetaobs+\thetajet)^3
    &\text{for } \thetajet < \thetaobs \\
  \end{cases}
\end{align}
These three regimes have different geometrical interpretations:
\begin{description}
  \item[$\thetaobs < 0$:]
    The observer is in the jet-opening angle, which means that  the jet looks like a spherical fireball. Most radiation comes from the angles close to the viewing axis.
  \item[$0 < \thetaobs \lesssim \thetajet$:]
    The observer is outside the opening angle, but only at a small angle from the jet edge. Therefore the jet geometry still contributes, and the observed flux is to be integrated over the observable part of the jet close to the edge.
  \item[$\thetajet < \thetaobs$:]
    The observer sees the fireball under a larger angle, so all regions of the fireball have approximately the same Doppler factor. Therefore the jet looks like a point source to the observer.
\end{description}

From these transformations it is clear that the jet looks different whether observed on- or off-axis, and we can derive relationships between quantities interpreted as off- versus on-axis observations; note that the on-axis quantities  are boosted into the shell rest frame with the Doppler factor $D(0)$. We define
\begin{align}
\label{equ:correction}
b \equiv 
\begin{cases} 
  \hphantom{a} \, D(0)/D(\thetaobs) &\text{for } 0 \le \thetaobs \lesssim \thetajet \\ 
  a \, D(0)/D(\thetaobs+\thetajet) &\text{for } \thetajet < \thetaobs \\ 
\end{cases}
\end{align}
with $a = D(2\thetajet)/D(\thetajet)$ chosen by demanding  that $b$ is continuous in $\thetaobs$. From the definition we note that always $ b\geq 1$. 
We can express the on-off-axis ratio for the isotropic equivalent energy as
\begin{align}
  \label{equ:on-off-transformation-eiso}
    \frac{E_\text{iso,on}}{E_\text{iso,off}} &= \begin{cases} b^2 &\text{for } \thetaobs < \thetajet \\ b^3 &\text{for } \thetajet < \thetaobs \\ \end{cases}\ ,
\end{align}
while the peak energy is simply Doppler shifted by
\begin{align}
 \frac{E_\text{peak,on}}{E_\text{peak,off}} = b \, .
\end{align}
Similar to $E_\text{iso}$, the variability timescale has to integrated over the shell geometry, as the radiation from different parts of the shell surface is delayed depending on $\thetaobs$. Following \cite{Ioka:2017nzl} we scale it as:
\begin{align}
  \label{equ:on-off-transformation-tvar}
  \frac{t_{v\text{,on}}}{t_{v\text{,off}}} & = \begin{cases} b^{-1} &\text{for } \thetaobs < \thetajet \\ b^{-1/2} &\text{for } \thetajet < \thetaobs \\ \end{cases}\ .
\end{align}
Note that \Ref~\cite{Monitor:2017mdv} implies that the duration of the burst $T_{90}$ scales with $b$ depending on the observation angle as well. However \Ref~\cite{Salafia:2016wru} argue that the observed burst duration does not change depending on the observation angle because it is defined in the source frame, which is at rest relative to the observer.
 
We do not re-scale $T_{90}$ with off-axis angle, which imples a larger number of interaction regions $N$ according to \equ{on-off-transformation-tvar} ($N \simeq T_{90}/t_v$) in the on-axis frame. While this may be counter-intuitive, the physical picture is that the many peaks in the lightcurve observed on-axis are smeared out off-axis, see discussion in \Ref~\cite{Salafia:2016wru}, leading effectively to a slower time variability if observed off-axis. In fact, the number $N$ drops out from the computation to a first approximation, and the smaller value of $t_v$ in the on-axis frame only slightly increases the neutrino production efficiency because the estimated shell width is derived from it. We have checked the impact of re-scaling $T_{90}$ as well, which however does not change the qualitative picture.

\subsection{Neutrino fluence for an off-axis observer}

The relationships in \equ{on-off-transformation-eiso}--\equ{on-off-transformation-tvar} describe the relation between on- and off-axis observables, leading to \eg\ lower energies for an off-axis observer if the on-axis observables are fixed.
Conversely, if the off-axis observables are fixed by observations, higher energies and shorter timescales are obtained in the on-axis frame. Note that the on-axis observables corresponds to the shell frame quantities Doppler-shifted by $D(0)$. 
While the secondary radiation calculated in the shell-frame has to be boosted back off-axis to predict observations, there is  still a net-effect. In the following analytical discussion, we focus on the case of small angles $\thetaobs < \thetajet$ for the off-axis transformations.

The $\gamma$-ray peak in the shell frame is shifted to higher energies as $E^\prime_{\gamma,\text{peak}} \propto b$, which implies that the neutrino production threshold is lower. As $E^\prime_{\nu,\text{peak}}\propto 1 / E^\prime_{\gamma,\text{peak}}$ (higher $\gamma$-energies lead to lower production thresholds), the observed neutrino spectrum will scale with $E_{\nu,\text{peak}} \propto b^{-2}$.

The neutrino production efficiency $f_\nu$ (the energy fraction the protons dumped into neutrino production) scales with the particle densities in the shell which depend on the luminosity and the dissipation radius. It can be estimated for the on-axis case from the pion production efficiency as~\cite{Waxman:1997ti,Guetta:2003wi}
\begin{align}
  \label{equ:pion-efficiency}
  f_\nu 
  \equiv \frac{E_{\nu,\text{iso}}}{\xi_A E_{\gamma,\text{iso}}} 
  \propto \frac{L_\gamma}{\Gamma^4 E_{\gamma,\text{peak}} t_v} \, ,
\end{align}
if the synchrotron cooling of the secondaries is neglected.

For small angles $\thetaobs \leq \thetajet$, the product $E_\text{peak} t_v$ is invariant under the observation angle and $f_\nu$ transforms proportional to luminosity $L_\gamma = E_{\gamma,\text{iso}} / t_\nu$ as $f_{\nu} \propto b^3/\Gamma^4$.

As the ratio $E_{\nu,\text{iso}} / E_{\gamma,\text{iso}}$ is invariant under observation angle while the neutrino peak is shifted by $\propto b^{-2}$,  an approximate scaling of the observed neutrino fluence $F_\nu$ in $[\text{GeV}^{-1}  \text{cm}^{-2}]$  can be obtained as
\begin{align}
  \label{equ:neutrino-fluence-scaling}
  F_{\nu, \text{off}}(E_{\nu}) \approx \frac{b^5}{\Gamma^4} F_{\nu, \text{on}}(b^{2} E_{\nu}) \, .
\end{align}
This means that the expected neutrino flux is enhanced when the observation is interpreted as off-axis- rather than on-axis-emission, while it is also shifted to lower energies.

\subsection{Photospheric constraint}

The scaling $E_\text{iso} \propto b^2$ has an additional implication for baryonically loaded jets. As the baryon density scales with the energy density, the shells are more opaque to $\gamma$-rays if observed off-axis with the same gamma-ray flux. 
The scaling can be read off from \equ{photospheric-radius}, where only $E_{\gamma,\text{iso,on}} \propto b^2$ is to be re-scaled:
\begin{align}
  \label{equ:photospheric-radius-scaling}
  &R_\text{ph} \simeq  \left(\frac{\sigma_T}{4 \pi m_p} \right)^{1/2} \, \left(\frac{\xi_A}{\varepsilon} \frac{E_{\gamma,\text{iso,off}}}{T_{90}/t_v} \right)^{1/2} \left(\frac{b^2}{\Gamma}\right)^{1/2} \\
  &= 5.4 \cdot 10^6 \text{km} \left(\frac{\xi_A}{10} \frac{t_v}{0.125\text{s}} \frac{E_{\gamma,\text{iso, off}}}{10^{47}\text{erg}}\frac{0.25}{\varepsilon} \frac{2\text{s}}{T_{90}} \frac{30}{\Gamma}\right)^{1/2} b
  \, . \nonumber
\end{align}
This condition effectively limits $\thetaobs$ and $\Gamma$ for a fixed baryonic loading if the emission ought to come from the dissipation from internal shocks beyond the photosphere. It can be  used to estimate the maximal allowed baryonic loading for which the shell is still transparent at $R_\text{coll}$ from the condition $R_{\text{coll}} \gtrsim R_{\text{ph}}$ as 
\begin{align}
  \label{equ:max-baryonic-loading}
  \frac{\xi_{A}}{\varepsilon} &\lesssim  \frac{4 \pi m_p }{\sigma_T}
  \, \frac{T_{90} t_v}{E_\text{iso, off}}
  \, \frac{4 \Gamma^5}{b^4} \\
  &=1.6 \cdot 10^{3} \frac{0.125\text{s}}{t_v} \frac{T_{90}}{2\text{s}} \frac{10^{47}\text{erg}}{E_\text{iso, off}} \frac{\Gamma^5}{30^5}\frac{1}{b^4} \nonumber
  \, .
\end{align}
For larger baryonic loadings, the radius where internal shocks develop will be below the photosphere -- where gamma-rays cannot escape. This is therefore the maximal baryonic loading allowed for the internal shock model. The neutrino flux computed for this value then also corresponds to the maximal allowed flux for this model.

\subsection{Numerical computation of neutrino fluence}

Our numerical calculations are based on the NeuCosmA tools~\cite{Hummer:2011ms}, see also \Refs~\cite{Boncioli:2016lkt,Biehl:2017zlw} for details. Isotopes are purely injected following a power law $\propto E^{-2} e^{-E/E_\text{max}}$. The injection luminosity is normalized to the energy density in gamma rays enhanced by the baryonic loading $\xi_A$. The isotopes will interact with the target photons of the SED described in Sec.~\ref{sec:observations}. Therefore, the neutrino fluence will mainly depend on the collision radius and the luminosity. Since for this event the observation is fixed, this translates into a dependence on the observation angle and Lorentz factor. 

Furthermore, we assume equipartition between magnetic and spectral energy density. The nuclei in the source are efficiently accelerated up to the maximum energy $E_\text{max}$, which is reached when the energy gain in the magnetic field is compensated by the loss processes. This has a mild impact on the neutrino fluence, since only the pion production threshold is relevant for the center-of-mass energy of nuclei interacting with photons. If in addition, the mean free path for photo-pion production is smaller than the size of the region, neutrino production will become efficient.

\section{Results}

\subsection{Structured (low luminosity) jet scenario}
\label{sec:structured-jet}

\begin{figure}[t!]
\centering
\includegraphics[width=0.45\textwidth]{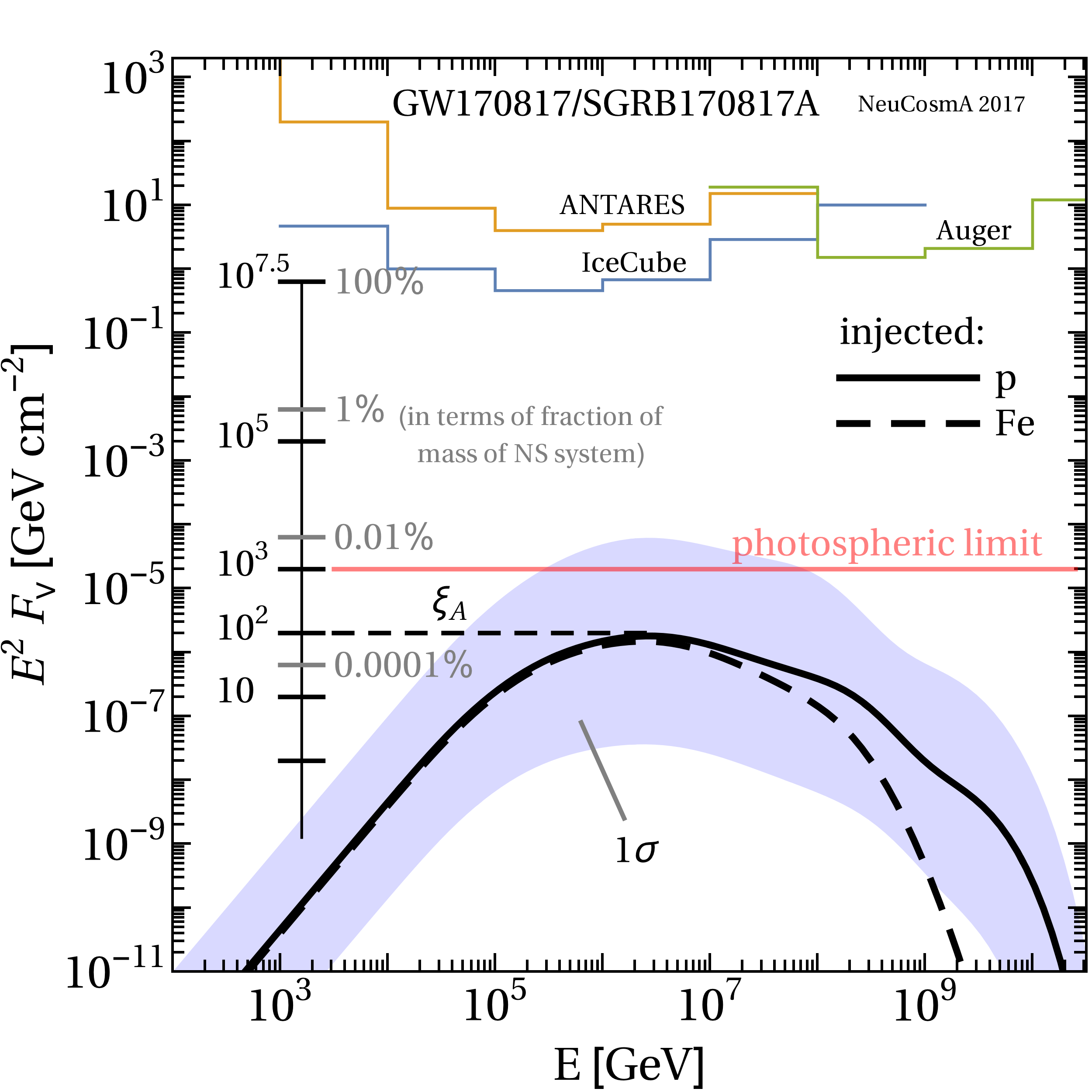}
\caption{Fluence of $\nu_\mu + \bar{\nu}_\mu$ for SGRB170817A assuming pure proton (solid) or iron (dashed) injections. The chosen parameters are $z = 0.008$, $L_X \sim 10^{47} \, \mathrm{erg \, s^{-1}}$, $t_v = 0.125 \, \mathrm{s}$, as given in \Refs~\cite{GBM:2017lvd,Goldstein:2017mmi}, and $\Gamma = 30$. The $1\sigma$-region includes the uncertainty of the measurement of these parameters given in the main text. The black scale indicates how the fluence will change with the baryonic loading $\xi_A$, with the gray percentage representing the fraction of the total mass of the NS merger. For a baryonic loading greater than $\xi_A = 10^3$, this collision would be sub-photospheric, indicated by the horizontal red line. Neutrino limits are taken from \Ref~\cite{ANTARES:2017bia}.}
\label{fig:nu-comp}
\end{figure}

In the first step, we assume a structured (low luminosity) jet as described above, where the Lorentz factor is fixed to $\Gamma = 30$. Fig.~\ref{fig:nu-comp} shows the predicted fluence of muon neutrinos for pure proton injection modeling SGRB170817A. As shown already in previous works \cite{Biehl:2017zlw,Biehl:2017hnb}, injecting nuclei heavier than protons shifts the cutoff of the neutrino fluence to lower energies, while there is only a slight impact on the peak for GRBs. 
This example has been computed with an initial baryonic loading of $\xi_A=100$, as indicated by the scale on the left side of the plot, it scales directly with this parameter. The blue band includes the $1\sigma$-uncertainties on the measured duration $T_{90}$, time variability $t_v$, redshift $z$, $\gamma$-ray fluence $F_\gamma$ as well as the spectral index $\alpha$ and peak energy $E_\text{peak}$ of the SED.
Note that we use $D=2 \Gamma$ instead of $\Gamma$ for the boost compared to what is frequently used in the literature.

The gray scale indicates which fraction of the total mass of the neutron star system has to be dumped into the jet. Assuming that the whole mass of the system, which is estimated to be $2.74^{+0.04}_{-0.01} M_\odot$ \cite{TheLIGOScientific:2017qsa}, goes into the jet, the maximum achievable baryonic loading is $\xi_A=10^{7.5}$. This is to be interpreted only as a rough guidance, since the actually realeased energy (compared to the isotropic equivalent energy) is smaller  by the beaming factor $\sim 1/(2 \Gamma^2)$ covered by the jet, which relaxes this constraint. On the other hand, for the structured jet scenario, the released energy in different directions may be higher, which makes the constraint stronger.

As an additional constraint, the photospheric radius scales with the baryonic loading. According to \equ{max-baryonic-loading} the maximum baryonic loading is $\xi_{A,\text{max}} \sim 10^3$ for the dissipation radius to be super-photospheric. This means that the shown neutrino fluence can be up-scaled by a factor of $10$ in this scenario, which represents our maximal possible neutrino fluence for this SGRB in the internal shock scenario.  Thus, if indeed neutrinos had been detected, then one would have concluded that the gamma-ray emission comes from the photosphere at a larger radius than the neutrino production radius.

\begin{figure}[t!]
\centering
\includegraphics[width=0.45\textwidth]{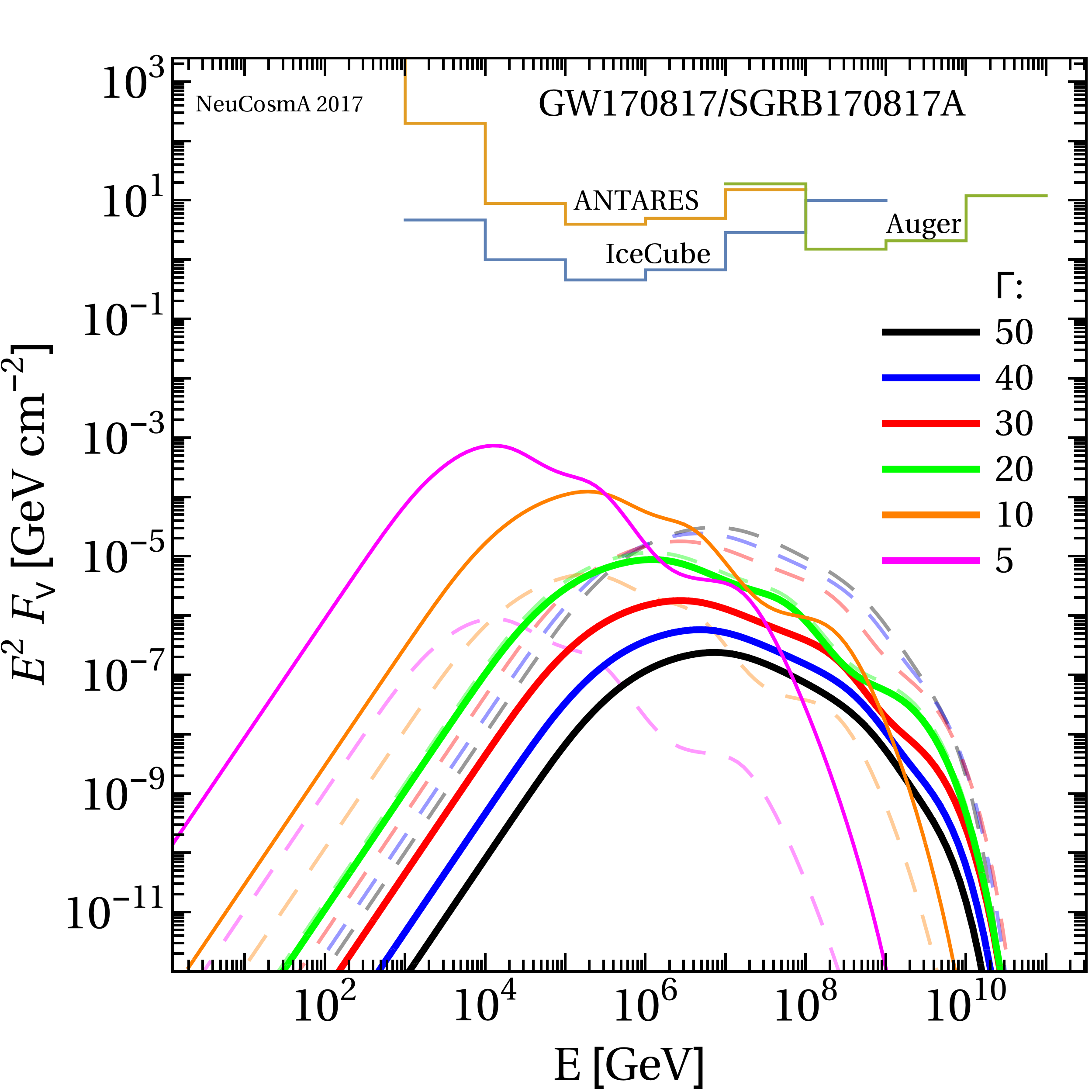}
\caption{Fluence of $\nu_\mu + \bar{\nu}_\mu$ for SGRB170817A and different values of the Lorentz factor $\Gamma$ in the structured jet case. We assume pure proton injection and the same parameters as given in Fig.~\ref{fig:nu-comp}. Solid curves refer to a fixed baryonic loading of $\xi_A = 100$, where thick solid curves correspond to collisions above the photosphere, and thin curves indicate sub-photospheric collisions. For the dashed curves, the baryonic loading has been maximized demanding that $R_{\mathrm{coll}}> R_{\mathrm{ph}}$.}
\label{fig:nu-gamma}
\end{figure}

We show the impact of the Lorentz factor on the muon neutrino fluence in Fig.~\ref{fig:nu-gamma}. 
The solid curves refer to a fixed baryonic loading $\xi_A = 100$, which illustrate that the fluence scales with $\Gamma$ according to \equ{neutrino-fluence-scaling} without imposing any additional constraints. The scaling agrees very well. However, for large shifts there is an additional damping of the high-energy tail of the spectrum due to secondary cooling, which was neglected in the simple analytic estimate \equ{neutrino-fluence-scaling}.

For low values of $\Gamma$, the collision radius decreases, which implies  efficient neutrino production. On the other hand, the photospheric radius increases, which leads to sub-photospheric collisions for $\Gamma \lesssim 20$ -- indicated by thin solid curves. The dashed curves indicate the maximal neutrino fluence using the photospheric constraint, which means that the curves for $\Gamma < 20$ are down-scaled to match it, and the curves for $\Gamma > 20$ are up-scaled accordingly. The expected maximal neutrino fluence is at most about four orders of magnitude below the neutrino telescope sensitivities, which means that the detection of a neutrino coming from this SGRB was extremely unlikely in the structured jet scenario.

\subsection{Off-axis fireball scenario}
\label{sec:off-axis-fireball}

\begin{figure}[t!]
\centering
\includegraphics[width=0.45\textwidth]{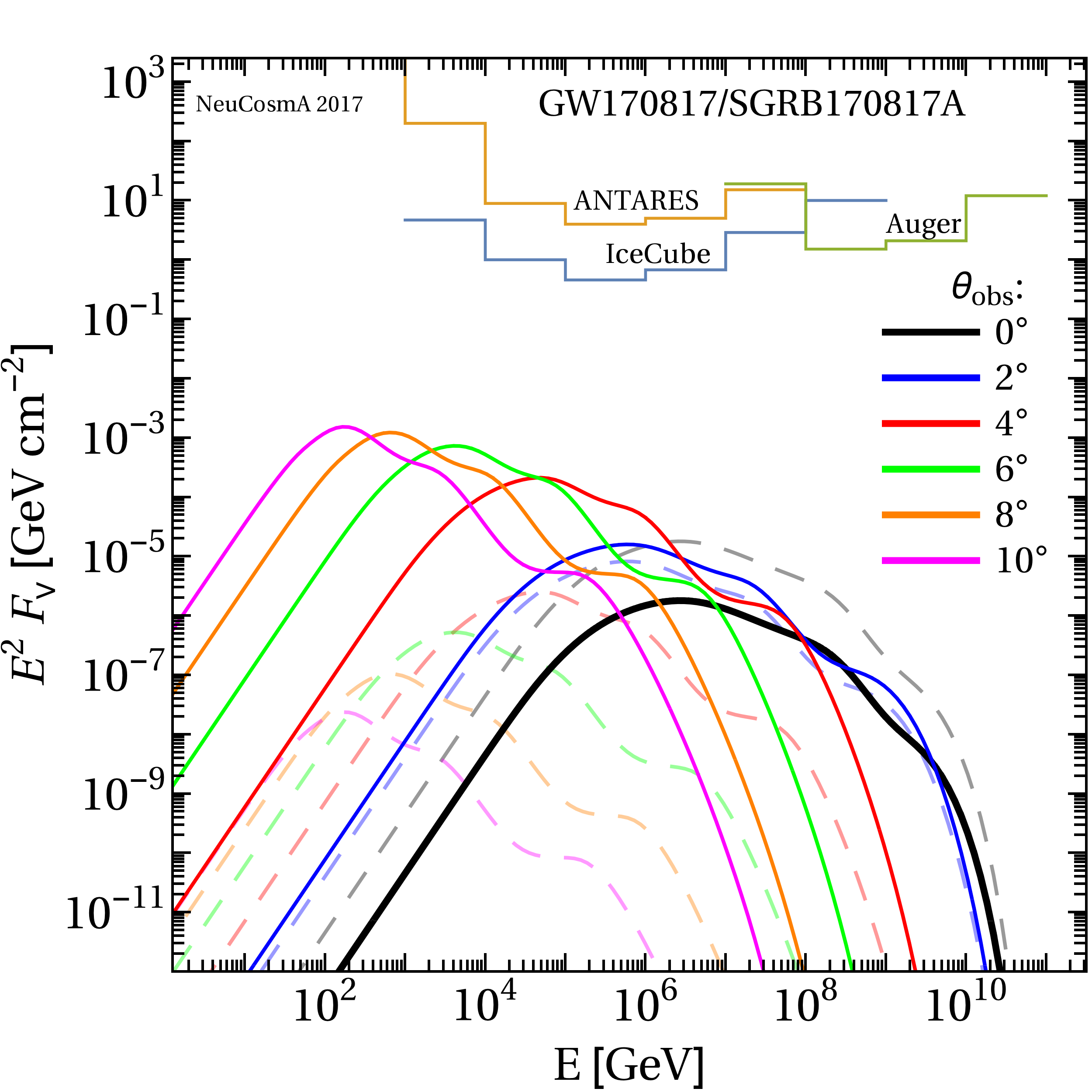}
\caption{Same as in Fig.~\ref{fig:nu-gamma} but for the off-axis (top hat) scenario, where we fixed the Lorentz factor to $\Gamma = 30$ and varied the observation angle $\theta_\text{obs}$.  Solid curves refer to a fixed baryonic loading of $\xi_A = 100$, where thick solid curves correspond to collisions above the photosphere, and thin curves indicate sub-photospheric collisions. For the dashed curves, the baryonic loading has been maximized demanding that $R_{\mathrm{coll}}> R_{\mathrm{ph}}$.}
\label{fig:nu-angle}
\end{figure}

In the off-axis fireball scenario, the observation angle $\thetaobs$ enters as an additional parameter influencing neutrino production and photospheric radius.

In Fig.~\ref{fig:nu-angle}, the dependence of the neutrino fluence on the observation angle is shown. The Lorentz factor is fixed to $\Gamma = 30$, which means that the scaling is given by \equ{neutrino-fluence-scaling}. Again, the solid curves represent the unscaled fluences with a fixed baryonic loading $\xi_A = 100$, while the dashed curves show the maximum achievable neutrino fluence corresponding to the solid curves re-scaled with the maximum possible baryonic loading demanding that $R_{\mathrm{coll}}> R_{\mathrm{ph}}$. From the way the curves rescale it can be deduced that the collisions become sub-photospheric (thin lines) already for small observation angles $\thetaobs \sim 2^\circ$ for this particular values of $\Gamma$ and $\xi_A$. For large observation angles, the fluence will be highly suppressed.
The maximum neutrino fluence is a few $\times 10^{-5}$ GeV cm$^{-2}$ for the on-axis observer and $\xi_{A,\text{max}} \approx 10^3$. Compared to the structured low luminosity jet, the off-axis observation makes it even less likely to detect a neutrino from this event.

\begin{figure}[t!]
\centering
\includegraphics[width=0.45\textwidth]{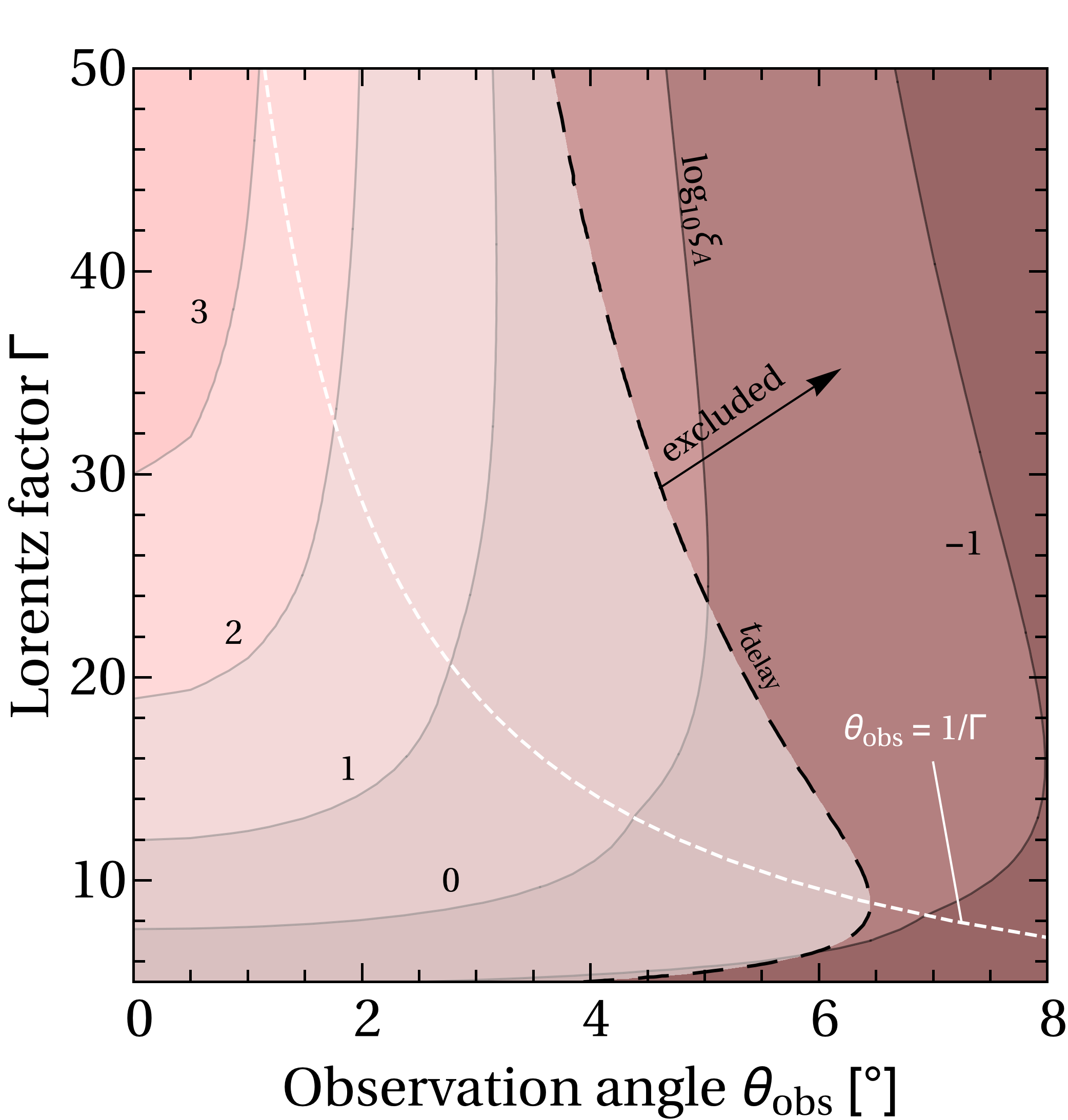}
\caption{Maximal baryonic loading $\xi_{A,\text{max}}$ such that the dissipation radius is still super-photospheric as a function of  $\thetaobs$ and Lorentz factor $\Gamma$.  It is assmed that $\thetajet \simeq 1/\Gamma$, which means that the scaling of the off-axis parameters changes for observation angles greater than $\theta_\text{obs} = 1/\Gamma$, indicated by the white dashed curve, according to \equ{on-off-transformation-eiso} and (\ref{equ:on-off-transformation-tvar}). The dashed black curve corresponds to a 1.7 s photon arrival time delay between gamma-ray and gravitational wave signal taken from \Ref~\cite{Salafia:2017hfr}. The dark shaded region would lead too larger arrival time delays, and is therefore excluded.}
\label{fig:scan}
\end{figure}

In order to demonstrate  how observation angle $\thetaobs$ and Lorentz factor $\Gamma$ are affected by the photospheric constraint, we show a parameter space scan in Fig.~\ref{fig:scan}. For each set of parameters, the maximum possible baryonic loading is calculated such that the collision is still super-photospheric in the internal shock scenario, indicated by the contours. We consider the change of scaling of the parameters \equ{correction} for large observation angles assuming that $\thetajet = 1/\Gamma$, \ie\ the white dashed curve indicates the break in the scaling for $\thetaobs = \thetajet$.

Additionally we show the constraint on these parameters by the measured time delay $t_\text{delay} = 1.7$ s between gravitational wave and electromagnetic signal given in \Ref~\cite{Salafia:2017hfr}, where the allowed region is highlighted in white.\footnote{Compared to the calculation given in \Ref~\cite{Salafia:2017hfr}, we assume a different efficiency $\varepsilon = 0.25$ to convert between the kinetic energy and the isotropic equivalent energy, leading to a slightly larger allowed region.} Note that we only use an upper limit on the time delay, since we do not assume that the shells are emitted at the same time as the gravitational waves -- they may be emitted later, which means that the delay could come from the engine. 
In fact, frequently used values for the baryonic loading in the literature are $\xi_A \sim 10$, for which the photospheric limit already provides stronger constraints: $\thetaobs \lesssim 3^\circ$ and $\Gamma \gtrsim 12$.

\begin{figure}[t!]
\centering
\includegraphics[width=0.45\textwidth]{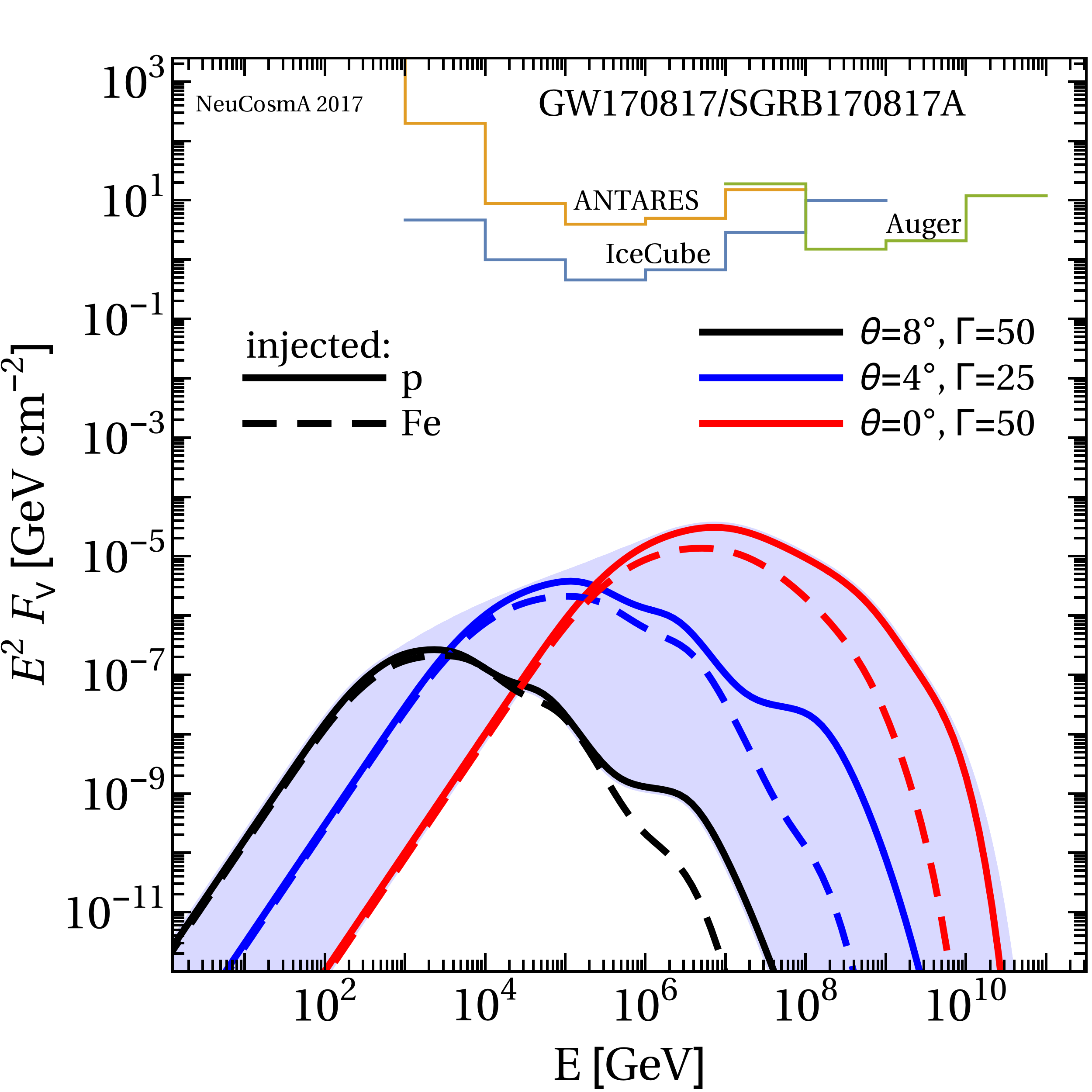}
\caption{Uncertainty on the fluence of $\nu_\mu + \bar{\nu}_\mu$ for SGRB170817A assuming pure proton (solid) or iron (dashed) injection and the same parameters as given in Fig.~\ref{fig:nu-comp}. The band includes the uncertainty of the observation angle $\thetaobs \in [0^\circ;8^\circ]$ and the Lorentz factor $\Gamma \in [5;50]$. Three example curves are shown with parameters generating a neutrino fluence covering different regions in the band (see legend). The individual curves have been re-scaled to the maximum possible baryonic loading satisfying $R_{\mathrm{coll}}> R_{\mathrm{ph}}$ for each set of parameters.}
\label{fig:uncertainty}
\end{figure}

In Fig.~\ref{fig:uncertainty} we show the uncertainty on the neutrino fluence obtained by varying observation angle and Lorentz factor in the range investigated in Fig.~\ref{fig:scan}. That is, for each point in the parameter space the fluence has been computed and rescaled with the maximal baryonic loading to obtain the maximum possible neutrino fluence. Any combination of parameters will generate a fluence within the blue uncertainty band. Most importantly, it will not exceed $\sim 5\times 10^{-5}$ GeV cm$^{-2}$, which is about a factor $10^{-4}$ below the sensitivity of the neutrino telescopes.
Thus, as it was possible to observe this event in $\gamma$-rays, it is clear that it is highly unlikely to see any neutrinos produced in the prompt phase of this SGRB in the internal shock scenario if the prompt and neutrino emissions come from the same dissipation radius.

We show several examples for combinations of $(\thetaobs,\Gamma)$ which occupy different regions of the allowed band. The higher the observation angle is, the more shifts the peak  to lower energies. In addition the supression is stronger since the photospheric constraint on these collisions is stronger. For the on-axis case, the peak is the highest, while intermediate values for $\thetaobs$ and $\Gamma$ produce neutrino fluences in between these two extremes. We show again solid curves for proton injection and dashed curves for the injection of $^{56}$Fe nuclei, which mainly results in an earlier cutoff of the neutrino spectrum.

Note again that rescaling with the maximum baryonic loading means that we do not allow for sub-photospheric collisions. If the photospheric constraint were omitted, \ie\ by performing a sub-photospheric extrapolation, the neutrino peaks would increase drastically. The reason is that for smaller radii the energy density would be much higher resulting in highly efficient neutrino production. However, Fig.~\ref{fig:nu-gamma} and~\ref{fig:nu-angle} indicate that even in the sub-photospheric extrapolation the peak is well below the sensitivity of the neutrino telescopes.

\section{Summary and conclusions}

We have computed the expected neutrino fluence from SGRB 170817A directly from the {\em Fermi}-GBM observation of the gamma-ray flux for two different scenarios: a structured jet scenario, and an off-axis top-hat scenario; see \figu{jetgeo}. In both cases,  the emission has been assumed to come from the dissipation in internal shocks, but under different off-axis angles. Consequently, we have also derived the constraint on off-axis angle and bulk Lorentz factor requiring that the emission region lies beyond the radius below which the source is optically thick to Thomson scattering. We have found that this constraint in fact limits the off-axis angle of the SGRB stronger than the delayed onset of the gamma-ray signal after the gravitational wave signal if the baryonic loading (energy in protons versus photons) is larger than ten. In that case, the off-axis angle (relative angle to edge of emission region) has to be smaller than about $3^\circ$, and $\Gamma \gtrsim 12$. Conversely, the baryonic loading, which the neutrino fluence is proportional to, is limited by this constraint. 

We have demonstrated that (without the photospheric constraint) the expected neutrino fluence strongly scales with $\Gamma$ and off-axis in a counter-intuitive way: the larger the off-axis angle, the higher the expected neutrino flux. The reason is that because we fix the observed gamma-ray fluence, the neutrino production efficiency will be larger for large off-axis angles due to higher radiation densities in the shell frame. However, we have also shown that the expected neutrino fluence can only be as high as $10^{-4}$ of the observed sensitivity if gamma-ray and neutrino emissions come from the same production region in the  internal shock scenario beyond the photospheric radius, even if the uncertainties on the GRB parameters including the baryonic loading are taken into account. 

We therefore conclude that no neutrinos from SGRB 170817A are expected from these theoretical models, in consistency with observations. The methods presented here can however be used to predict the neutrino fluence from future GRBs, if observed off-axis, and also show some of the subtleties which need to be observed in off-axis computations.

\subsection*{Acknowledgments}

We thank Denise Boncioli for useful discussions. 
This work has been supported by the European Research Council (ERC) under the European Union’s Horizon 2020 research and innovation programme (Grant No. 646623).

\bibliographystyle{apsrev4-1}
\bibliography{references}

\end{document}